\documentclass[sigchi]{acmart}

\usepackage{booktabs} 
\usepackage{balance}       

\setcopyright{rightsretained}

\copyrightyear{2019} 
\acmYear{2019} 
\setcopyright{iw3c2w3} 
\acmConference[CHI 2019]{CHI Conference on Human Factors in Computing Systems Proceedings}{May 4--9, 2019}{Glasgow, Scotland Uk}
\acmBooktitle{CHI Conference on Human Factors in Computing Systems Proceedings (CHI 2019), May 4--9, 2019, Glasgow, Scotland Uk}
\acmDOI{10.1145/3290605.3300690}
\acmISBN{978-1-4503-5970-2/19/05}

\begin{document}
\title{Gamification in Science: A Study of Requirements in the Context of Reproducible Research}

\author{Sebastian S. Feger}
\affiliation{%
  \institution{LMU Munich and CERN}
  \city{Munich, Germany and Geneva, Switzerland}
}
\email{sebastian.s.feger@cern.ch}

\author{S{\"u}nje Dallmeier-Tiessen}
\affiliation{%
  \institution{CERN}
  \city{Geneva}
  \state{Switzerland}
}
\email{sunje.dallmeier-tiessen@cern.ch}

\author{Pawe{\l} W. Wo{\'z}niak}
\affiliation{%
  \institution{Utrecht University}
  \city{Utrecht}
  \country{the Netherlands}
}
\email{p.w.wozniak@uu.nl}

\author{Albrecht Schmidt}
\affiliation{%
  \institution{LMU Munich}
  \city{Munich}
  \country{Germany}}
\email{albrecht.schmidt@ifi.lmu.de}

\renewcommand{\shortauthors}{S. Feger et al.}

\begin{abstract}
The need for data preservation and reproducible research is widely recognized in the scientific community. Yet, researchers often struggle to find the motivation to contribute to data repositories and to use tools that foster reproducibility. In this paper, we explore possible uses of gamification to support reproducible practices in High Energy Physics. To understand how gamification can be effective in research tools, we participated in a workshop and performed interviews with data analysts. We then designed two interactive prototypes of a research preservation service that use contrasting gamification strategies. The evaluation of the prototypes showed that gamification needs to address core scientific challenges, in particular the fair reflection of quality and individual contribution. Through thematic analysis, we identified four themes which describe perceptions and requirements of gamification in research: Contribution, Metrics, Applications and Scientific practice. Based on these, we discuss design implications for gamification in science.
\end{abstract}

%
%



\maketitle

\section{Introduction}
Reproducibility should be a scientific cornerstone that enables knowledge transfer and independent research verification. Yet, alarming reports describe the systematic failure to reproduce empirical studies in a variety of scientific fields ~\cite{Baker2016}. \textit{Preserving} and \textit{sharing} research are key reproducible practices ~\cite{Bechhofer2013, Wilkinson2016}, which require efforts to prepare and document experimental resources ~\cite{Borgman:1297241}. But those efforts are often not matched by perceived gains ~\cite{Begley2012,Collaboration2012}.

Gamification, the "use of game design elements in non-game contexts" \cite{Deterding2011}, has proven to be a valuable tool for engaging users and motivating desired behaviors \cite{Cavusoglu2015}. In this paper, we explore possible uses of gamification to support reproducible practices. Past efforts attempted to stimulate good scientific practices through open science badges. They have been shown to significantly impact research data sharing practices of publications in the \textit{Psychological Science} journal \cite{kidwell2016badges}. Yet, little empirical knowledge exists on the moderating effects of individual gamification mechanisms in professional scientific settings. In fact, it has recently been argued that mapping the impact of game design elements on specific domains is indeed much needed in gamification research today \cite{Nacke2017}.

We developed and evaluated two gamification prototypes that are inspired by an existing High Energy Physics (HEP) research preservation service. Both aim to stimulate research documentation and sharing. A series of research activities informed the designs: we reviewed field studies, conducted interviews and observed a workshop, to learn about physicists' needs and perceptions towards the service. This approach is in line with increasing evidence that meaningful gamification profits from in-depth knowledge of target users \cite{Brito2015, Dale2014}, and calls to consider unique characteristics and frameworks of scientists in scientific gamification design \cite{feger2018just}. 

Our paper makes four contributions. First, we reinforce the need for systematic, user-centered gamification design and detail our researcher-centered design process. Second, we contrast two gamification designs that aim to encourage key practices of reproducible research, one of the most prevalent challenges in science. Third, we compare the perceived value, enjoyment, suitability and persuasiveness of the two gamification strategies and provide qualitative insights to explain why distinct mechanics motivate or alienate scientists. Finally, we present design implications for gamification in scientific environments and discuss how those relate to open science badges, that have shown to impact researchers' sharing behaviors.

This paper is organized as follows: We first reflect on needs of meaningful gamification and highlight how our study addresses emerging challenges of gamification research. Next, we provide details of the context and the study design. We further detail the design process of the two interactive prototypes and present the results and findings of their evaluation. Finally, we discuss design implications for gamification in science.

\section{Related Work}
In this section, we reflect on past gamification design and moderating effects of game mechanisms. We further depict how this study relates to emerging research challenges, gamification in science and reproducible research.

\subsection{Meaningful Gamification}
The current iteration of game-inspired motivational design is characterized as the "use of \textit{game elements} in non-game contexts" \cite{Deterding2011}. In their work, Hamari et al. \cite{hamari2014does} conducted a literature review of empirical studies on gamification. They report on 10 investigated motivational affordance categories, including levels, stories, clear goals, feedback, rewards, progress and challenges, noting that "points, leaderboards, and badges were clearly the most commonly found variants." As becomes increasingly evident, gamification that adds leaderboards, points and badges only to drive business goals, is likely to prevent long-lasting engagement and even risks alienating users \cite{nicholson2015recipe}. Instead, gamification profits from a holistic design process that appeals to the intrinsic motivation of the players \cite{Brito2015,Dale2014}, requiring systematic, user-centered designs \cite{werbach2012win, kumar2013gamification}. This motivated our researcher-centered gamification design process. Intrinsic motivation results from activities that are perceived as satisfactorily or pleasurably in themselves. Self-determination theory (SDT), as described by Ryan and Deci \cite{ryan2000self}, distinguishes intrinsic and extrinsic motivation. Extrinsic motivation is based on external goals that represent outcomes separable from the activity itself (e.g. rewards, money or social approval) and intrinsic motivation comes from psychological needs: Competence, Autonomy and Relatedness.

\subsection{Mapping Impact of Game Design Elements}
In \textit{Rethinking Gamification} \cite{fuchs2014rethinking}, Deterding stresses that "motivational design should revolve around designing whole systems for motivational affordances, not adding elements with presumed-determined motivational effects." In \textit{The Maturing of Gamification Research}, recently published by Nacke and Deterding \cite{Nacke2017}, the authors highlight that gamification's early research focused on few contexts, like education. As not all contexts and desired behaviors are equally suited for gamification, "extending the use of gamification beyond these contexts, and systematically studying the moderating effects of different individual and situational contexts is thus very much in need today". The authors argue that "we are just at the beginning of understanding what gamification design elements and methods best map onto what application domains". Recent work from Orji, Tondello and Nacke \cite{orji2018personalizing} represents a good example, mapping impact of persuasive strategies on gamification user types for persuasive gameful health systems. Basing their study on storyboards, they illustrate how gamification research profits from novel methods. This approach also inspired our prototype-centered evaluation, mapping moderating effects of game design elements in science.

\subsection{Gamification in Science}
Studying gamification in a research setting represents an opportunity to extend our knowledge of the applicability and constraints of gamification beyond traditional contexts. So far, gamification in science focused on designing engaging experiences in citizen science, motivating the general public to contribute to scientific knowledge through micro tasks \cite{Eveleigh2013,Bowser2014}, and supporting the learning process of students \cite{Ibanez2014}. CHI workshop summary from Deterding et al. \cite{deterding2015gamifying} raises questions on the role of gamification in research, but also focuses on citizen science, encouraging users to provide self-tracking data and to participate in research activities.

In their recent work, \textit{Just Not The Usual Workplace: Meaningful Gamification in Science}, Feger et al. \cite{feger2018just} motivate the need for research on the role of gamification in scientific work environments. They highlight that scientists' needs, practices, motivations and socio-technical frameworks differ from corporate employees and are likely to impact their perceptions of gamification. Having little to no empirical understanding of the needs and constraints of gamification in this environment, gamification research needs to consider "scientific environments as a special type of workplace and scientists as unique type of employee." Our researcher-centered gamification design process is also inspired by the authors' call to systematically study target users. They propose to review existing social scientists' studies on field practices, field differences and scholarly communication in the target field as part of a systematic design process.

\subsection{Gamification for Reproducible Research}

The reproducibility crisis represents a strong example of scientific challenges that motivate studying needs and constraints of gamification in research settings. Documenting and sharing research data and resources are key requirements of reproducible research \cite{Bechhofer2013, Wilkinson2016}. But, the efforts required to prepare, document and share experimental data ~\cite{Borgman:1297241} are often not matched by the perceived gain. In fact, studies claim that the scientific culture does not support or even impairs compliance with reproducible practices \cite{Begley2012,Collaboration2012}. Considering missing incentive structures, common proposals target the implementation of policies; most prominently concerning funding rules ~\cite{russell2013if} and requirements for conference and journal submissions ~\cite{Belhajjame2014,Stodden2014}. 

Persuasive gamification design might provide further motivation for scientists to conduct reproducible research. Kidwell et al. \cite{kidwell2016badges} report on the impact of the \textit{Psychological Science} journal adopting open science badges. Authors who made their data and / or materials openly available, received corresponding badges, displayed on top of their paper. In their quantitative study, they find that badges increased reported and actual sharing rates significantly, both in comparison to previous sharing behaviors in the same journal and other journals in the same discipline. Yet, despite this indication that game elements can significantly impact open sharing practices, empirical studies on the moderating effects of gamification in science are still missing.

\section{Method}
\begin{figure}
  \centering
  \includegraphics[width=1.0\columnwidth]{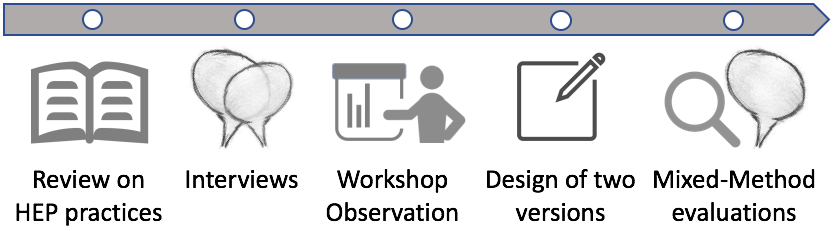}
  \Description{Details our research process. First, we reviewed High Energy Physics practices. Second, we conducted interviews. Third, we observed a workshop. Next, we designed the two prototypes. Finally, we conducted a mixed-method evaluation of those prototypes.}
  \caption{Our design process focused on practices in HEP and perceptions towards a preservation service. Two distinct designs were evaluated with experimental physicists.}~\label{fig:research_process}
\end{figure}

This section details the study context and various research activities, depicted in Figure \ref{fig:research_process}, conducted during the design and evaluation process.

\subsection{Study Context}

We conducted our study with data analysts working for the major experiments of the Large Hadron Collider (LHC), the world's biggest particle accelerator. The LHC is located at CERN, a key High Energy Physics (HEP) laboratory. CERN and HEP provide an optimal context for studying the challenges of research preservation and reproducibility, as they represent one of the most data-intensive research environments in the world \cite{Mélissa:2276551}. The community stands out in the early adoption of computer-supported technology, most notably the World Wide Web \cite{Berners-Lee1992,birth:1998446}, designed to bridge large, dislocated research collaborations. Today, thousands of physics researchers are working in the biggest LHC collaborations ALICE, ATLAS, CMS and LHCb. 

CERN Analysis Preservation (CAP)\footnote{Publicly available on GitHub: 

\url{https://github.com/cernanalysispreservation}} ~\cite{chen2016cern}, currently a prototype service, enables researchers to fully describe their analyses, consisting of data, metadata, workflows and code files ~\cite{Cowton2015}. CAP offers a web-based graphical user interface and a command-line client that support physicists in documenting and preserving their analyses. Due to differing structures, each analysis description form is tailored to the experiment to which it belongs. Initially, analyses on CAP are visible only to the creator. Once submitted, analyses can be shared with the whole collaboration or individual collaboration members.

\subsection{Gamification Designs}
In line with existing gamification design models that emphasize studying needs, practices and motivations of target users, we set out to learn about HEP data analysts. As it has been proposed \cite{feger2018just}, we started by reviewing published studies on field practices, field differences and scholarly communication in HEP. Next, we conducted semi-structured interviews with 12 HEP data analysts to learn about their workflow practices and perceptions of contributing to CAP. We were interested in learning about: current preservation and information seeking practices; expected benefits resulting from a preservation service; and perceptions towards a preservation service. We collected 9 hours of recordings that were transcribed and coded. Finally, we observed a one-day workshop that was attended by representatives of the four biggest LHC collaborations. The service developers presented the latest features and collaboration representatives discussed their needs and wishes for the future service development. We gained full access to the workshop notes and presentations.

To stimulate feedback, we created two designs that are based on our researcher- and service-centered insights. Following our initial expectation that gamification in a professional scientific context is most likely to profit from a serious, informative and rule-based design language, we created the Rational-Informative Design (RID) that provides information expected to appeal to HEP researchers. The RID was designed to make less use of most common game elements like points and leaderboards. Instead, it uses elements of "Social networks", "Social discovery", "Signposting" and "Challenges" (the LHC collaboration goal on the dashboard) as suggested by Tondello et al. \cite{Tondello2017}. This enables an exploration of gameful design elements in the HEP context. Yet, as scientists are already subjected to a high degree of competition, we also created a contrasting Simple Game Elements Design (SGED) version that focuses on point-based rewards and competitive elements. The basic UI design rules (color scheme, arrangements, etc.) are the same for both versions and are inspired by the actual service design. We built interactive wireframes with a high level of detail using the prototyping tool Balsamiq. As it is impractical to develop two functional, productive designs for the purpose of studying perceptions, we decided in favor of fully interactive prototypes. This approach is also motivated by recent, novel research methods, mapping persuasive strategies to gamification user types based on storyboards \cite{orji2018personalizing}. 

\subsection{Evaluation}
We conducted mixed-method, within-subjects evaluations with 10 HEP data analysts. As indicated in Table \ref{tab:table_participants}, researchers from CMS and LHCb were primarily recruited, as their CAP submission forms are most complex and time demanding. We particularly considered recruiting participants with a very diverse set of professional experiences and roles. Participants who had finished their PhD three or more years ago were considered \textit{senior.} We further identified current or previous \textit{conveners}, physicists who have a particular project management role within a collaboration. The 10 participants included 3 female researchers, reflecting the employment structure of research physicists at CERN \cite{CERN-HR-STAFF-STAT-2017}. The analysts' ages ranged from 27 to 53 years old (Avg = 36, SD = 8.2). No remuneration was provided, as all evaluation sessions were conducted during normal working hours and all participants were employed by CERN or an associated institute.

\begin{table}
  \centering
  \begin{tabular}{l c c c r}
    {\small\textit{Ref}}
    & {\small \textit{Affiliation}}
    & {\small \textit{Gender}}
    & {\small \textit{Experience}}
    & {\small \textit{Order}}\\
    \midrule
    P1 & CMS & Male & Senior & SGED-RID \\ 
    P2 & LHCb & Male & Postdoc & RID-SGED \\ 
    P3 & CMS & Female & Senior & SGED-RID \\
    P4 & LHCb & Female & PhD student & RID-SGED \\
    P5 & CMS & Female & Senior & SGED-RID \\
    P6 & LHCb & Male & PhD student & RID-SGED \\
    P7 & CMS & Male & Senior, Convener & SGED-RID \\
    P8 & ATLAS & Male & Senior, Professor & RID-SGED \\
    P9 & CMS & Male & Senior & SGED-RID \\
    P10 & CMS & Male & Postdoc & RID-SGED \\

  \end{tabular}
  \caption{Researchers with a diverse level of experiences and roles were recruited.}~\label{tab:table_participants}
\end{table}

\subsubsection{Structure}
First, participants were quickly introduced to CAP and were shown the analysis submission form of their corresponding collaboration, in order to get familiar with the context. Afterwards, half of the participants started exploring the RID version, the other half the SGED one. They started with the dashboard and explored the various views on their own. We prepared a few questions for every principal view that aimed to stimulate feedback.
Following the design exploration, we asked the physicists to respond to a 7-point Likert scale questionnaire, structured as follows:

\begin{itemize}
    \item An abbreviated Intrinsic Motivation Inventory (IMI) scale was used, as it provides two valuable subscales. We considered assessing the perceived \textbf{Value / Usefulness} (5 items) to be of key importance for gamification in science, as well as \textbf{Interest / Enjoyment} (4 items). Enjoyment has also been used to characterize user preferences of game design elements by Tondello et al. \cite{Tondello2017}. The interest / enjoyment subscale assesses intrinsic motivation per se, while task meaningfulness appeals to the innate need for autonomy \cite{sailer2017gamification}.
    \item We further asked to rate a statement that targets the \textbf{suitability} of the design: \textit{The system is NOT suitable for a research preservation service}. Finally, \textit{The system would influence me to document my analyses}, targets the \textbf{persuasiveness} of the design, also core to the study of Orji et al. \cite{orji2018personalizing}.
\end{itemize}

Afterwards, participants explored the other version and the process was repeated. In the following, we asked analysts to compare the two versions. Finally, analysts were invited to fill in a short questionnaire with six items, assessing the validity of our underlying design assumptions.

\subsubsection{Data Analysis}
We collected 5.2 hours of recording during the evaluation sessions. All recordings were transcribed non-verbatim and Atlas.ti data analysis software was used to organize, analyze and code the transcriptions. We performed Thematic Analysis \cite{Blandford:2222613} to identify emerging themes. Two authors independently performed open coding of the first two transcriptions. They discussed and merged their codes. The resulting code tree was used in coding the remaining transcriptions. A hundred and three codes and 10 code groups resulted from this combined effort. The code tree was used as reference in coding the remaining transcriptions. In total, 124 codes were created through 287 quotations (1 - n associated codes). Code groups were adapted and merged, resulting in 9 code groups. We constructed the four high-level themes based on those code groups. For example, the theme "Scientific practice" is based on "Speaking scientists' language" and "Known mechanisms".

\section{Design}
In this section, we report on our researcher-centered design process and present the interactive prototypes of a gamified research preservation service.

\subsection{Researcher-Centered Design}
Here, we first detail the insights gathered from our research activities, studying practices and motivations of HEP data analysts, as well as perceptions towards research preservation. Based on those, we present target behaviors for the gamification design.

\subsubsection{Literature Review: HEP Field Practices}
Various studies report on the role researchers play within the huge collaborations. In her article called "The Large Human Collider" \cite{Merali2010}, Merali documents the high level of identification with the detector. She devotes an entire section to researchers sacrificing their identity to their respective LHC collaboration. Merali refers to Karin Knorr Cetina, a sociologist who has been studying CERN's collaborations for almost three decades. Knorr Cetina confirms that CERN "functions as a commune, where particle physicists gladly leave their homes and give up their individuality to work for the greater whole." In her earlier work, she even described "the erasure of the individual epistemic subject in HEP experiments." \cite{cetina2009epistemic} 

\subsubsection{Interview: Practices, Needs and Perceptions}

In our interviews, participants reported commonly sharing their analysis resources (codes, datasets and configurations) with their colleagues. Yet, we realized that despite the very early invention and adoption of collaborative technologies, the information and communication architecture is shaped by traditional forms of communication. Searching for resources is hindered by challenges imposed by the databases or unstructured presentation of materials. As analysts are in high demand of information, they still rely heavily on E-Mail communication, trying to satisfy their information and resource needs through personal networks.

The communication architecture results in a high level of uncertainty. Participants highlighted that reliably informing all dependent analysts about discovering an issue in a common analysis resource is difficult, if not impossible. E-Mail communication in collaborations with several thousand members and highly distributed institutes is not sufficient. Thus, we envision documentation as a strategy to cope with uncertainty, as the service could reliably inform every analyst who depends on a marked resource.

The interviews revealed the value of collaboration in HEP, as well as challenges of engaging in collaborative behavior, imposed by the information architecture. Analysts cannot know all relevant colleagues in their highly distributed collaborations. We envision rich analysis documentations as a strategy to increase the visibility of researchers, thereby improving chances to engage in collaboration.

Finally, analysts reinforced the value of centralized, automated execution. They highlighted the efforts of setting up their own environments and acquiring computing time. Some of the analysts run their analyses on their institute's servers, as there is less competition for computing resources. However, doing so hinders sharing and collaboration with researchers outside their institute and requires substantial efforts when changing institutes. Thus, automated analysis re-execution on a centralized preservation service represents a strong incentive to keep documented analyses up-to-date.

\subsubsection{Workshop Observation}
The service developers presented for the first time the full-cycle execution from CAP to REANA\footnote{http://www.reana.io/}, a partner project that aims to re-execute analysis workflows on a centralized computing framework. Matching our interview analysis, this functionality was acknowledged very positively by the attending researchers. It confirmed our initial thoughts of promoting execution of appropriately documented and structured analyses on REANA. As we learned in the workshop, there is a second dimension to analysis structuring and automation, which relates to the use of workflow management systems. Making use of such tools fosters scalable, machine-readable and executable analysis designs, representing an important step towards automated re-execution. 

\begin{figure*}
  \centering
  \includegraphics[width=2.1\columnwidth]{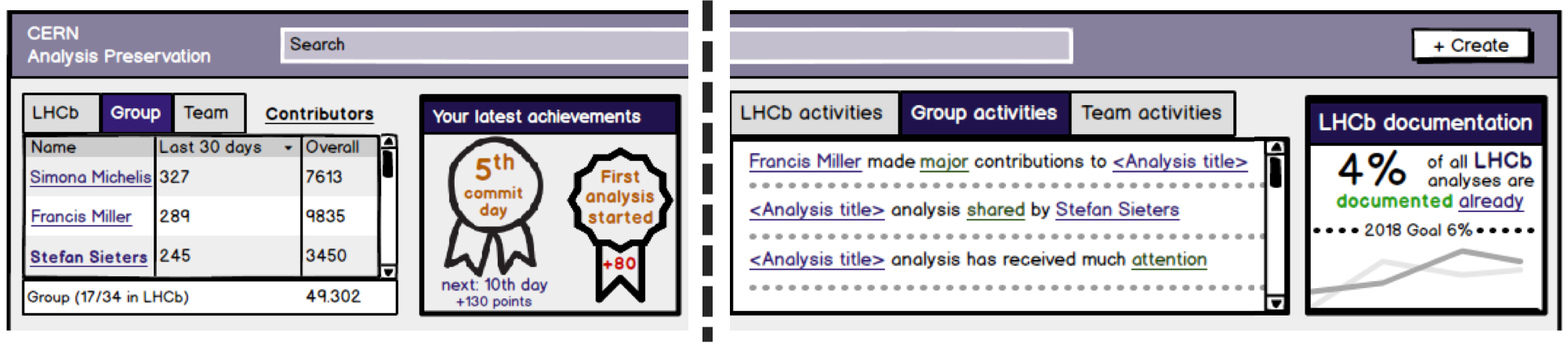}
  \Description{Shows excerpts of the dashboard from both prototypes. The SGED dashboard shows a container called "Your latest achievements", displaying two badges. It also shows a leaderboard featuring contributors. The RID dashboard shows a activity list and statistics about the development of contributions to the corresponding LHC collaboration.}
  \caption{The dashboards make use of contrasting mechanisms. While the SGED (left) shows latest achievements, rewards and point-based leaderboards, the RID (right) shows collaboration-wide documentation statistics and a rule-based activity stream.}~\label{fig:figure_dashboard_combined}
\end{figure*}

\begin{figure*}
  \centering
  \includegraphics[width=1.7\columnwidth]{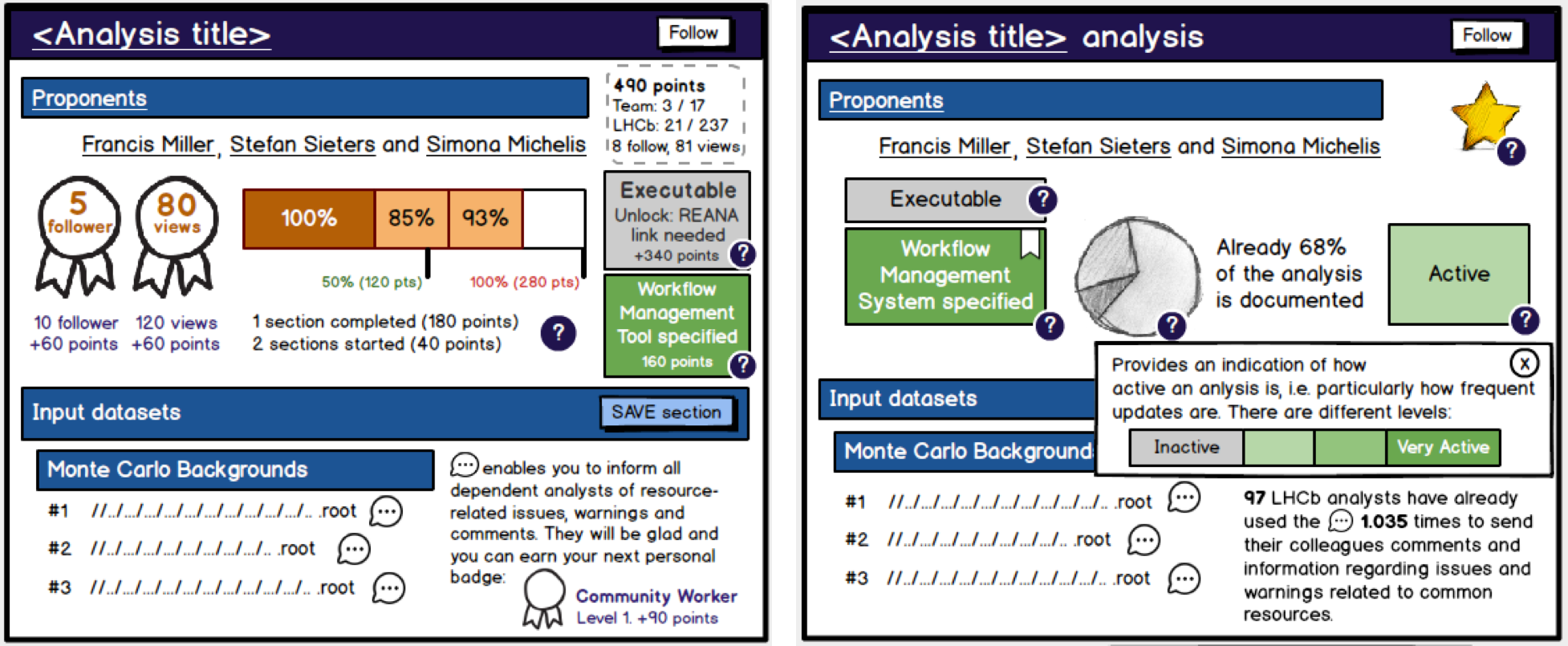}
  \Description{The image shows analysis pages from both versions next to each other. The SGED version has badges, progress bars and challenges with indicated point-based rewards. The RID analysis page avoids the use of points, but rather reflects an analysis' status rationally.}
  \caption{The analyses pages of the two prototypes. While the SGED (left) analysis page promotes point-based rewards, the RID analysis page reflects an analysis status.}~\label{fig:analysis_pages}
\end{figure*}

\subsubsection{Target Behaviors (TB)}
Based on previously described research activities, we developed four target behaviors that we want to encourage through our designs:

\begin{itemize}
    \item \textbf{TB\#1 - Document and provide rich descriptions. } Primarily, we want to encourage data analysts to document and preserve their work thoroughly.
    \item \textbf{TB\#2 - Communicate relevant information. } Analysts discovering issues related to collaboration-shared analysis fragments should be encouraged to share their concerns. In turn, we expect to create awareness that documenting dependencies can be a strategy to cope with uncertainty.
    \item \textbf{TB\#3 - Use automation capabilities. } Structuring and documenting analyses for central re-execution represents an opportunity to speed up analysis workflows. We expect physicists who follow this TB to experience benefits through automated, more efficient workflows that, in turn, provide motivation to keep the documentation up-to-date.
    \item \textbf{TB\#4 - Embrace collaborative opportunities. } A central research preservation service provides opportunities for analysts to increase their visibility; and the visibility of their work - likely leading to valuable collaboration. Stimulating researchers to pursue high visibility, we expect analysts to document and share their work.
\end{itemize}

\subsection{Prototypes}

Two interactive prototypes of a gamified research preservation service were developed. They are based on two principal views: dashboard and analysis page. They are inspired by views that already exist in CAP. In addition, we created a profile page to list activities and achievements. In this section, we depict the dashboard and analysis pages of both versions. The complete, interactive, high-resolution Balsamiq archives are provided as supplementary material.

\subsubsection{Simple Game Elements Design (SGED)}
As shown in Figure \ref{fig:figure_dashboard_combined}, the dashboard design of the SGED (left) focuses on achievements and competition. Latest rewards are displayed in the center, together with awarded points and upcoming challenges. As indicated by the shown badges, they primarily attempt to stimulate documentation (TB\#1). Two leaderboards depict the performance of contributors and the relevance of analyses. In order to foster collaboration (TB\#4), leaderboards can be set to show the whole collaboration or the single group or team. Listed contributors link to corresponding profile pages and analysis titles to analysis pages. 

The analysis page, depicted in Figure \ref{fig:analysis_pages} (left), educates and stimulates researchers towards using central computing resources for automated (re-)execution of analyses (TB\#3). Badges are awarded both for establishing a link with the REANA project and the integration of a workflow management system. Analysis points are awarded to the analysis, as well as to all proponents. Having learned about the importance of visibility and collaboration (TB\#4), we added rewards and challenges that target analysis impact (views / followers). The documentation progress bar gives a visible overview of the completeness of the analysis and incentivizes further contributions (TB\#1). Finally, the importance of sending relevant resource-related information is highlighted; and compliance incentivized (TB\#2).

\subsubsection{Rational-Informative Design (RID)}

The dashboard in this version displays an activity stream. As depicted in Figure \ref{fig:figure_dashboard_combined} (right), researchers can as well control the desired granularity (TB\#4). Entries in the stream refer to a researcher and / or analysis and a \textit{rule} that needs to be fulfilled (TB\#1). When clicked, further information concerning a rule are shown; as well as analyses that comply with it (TB\#4). Having learned about the particularly strong identification of HEP researchers with their collaboration, we decided to depict the collaboration's preservation status and a community goal. Thereby, we envision triggering researchers' sense of identification to stimulate contributions (TB\#1) that impact a common good of the collaboration.

The analysis page, shown in Figure \ref{fig:analysis_pages} (right), is designed to rationally report on statuses and does not make use of point-based rewards. Badges are used to educate and indicate use of automated, centralized analysis (re-)execution and workflow tools (TB\#3). A pie chart indicates the number of blocks that are fully, partially or not at all documented. Depending on the level of documentation, encouraging messages are shown (TB\#1). Analyses that continue to receive contributions (TB\#1) are indicated as \textit{active}. 
Based on our previous research, we expect this to be a meaningful attribution, as active analyses are more likely to be of interest to other collaboration members (TB\#4). A star marks popular analyses that have many followers and views. Finally, information is provided on the usage of the resource-related communication feature (TB\#2). Detailing the number of analysts who have used the feature, we aim to stimulate the identification of analysts with their collaboration and provide an opportunity to directly impact those collaboration-related statistics.

\section{Results}

\begin{figure}
  \centering
  \includegraphics[width=0.9\columnwidth]{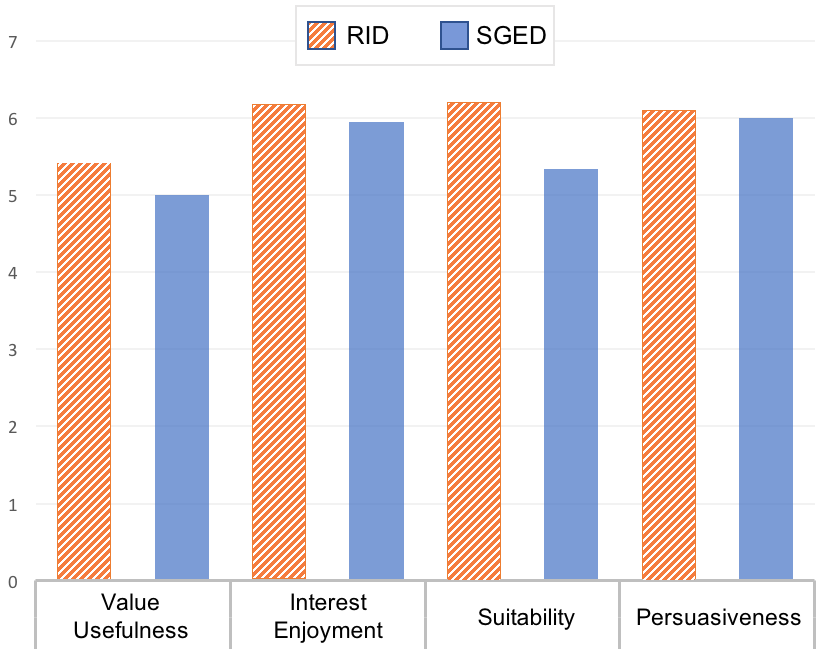}
  \Description{Depicts the results from the quantitative evaluation. The results are as follows [Mean (SD)]. \textit{RID}: Value $5.42$ ($0.95$), Interest $6.18$ ($0.75$), Suitability $6.2$ ($0.75$), Persuasiveness $6.1$ ($1.04$). \textit{SGED: Value $5.0$ ($0.45$), Interest $5.95$ ($1.06$), Suitability $5.3$ ($1.55$), Persuasiveness $6.0$ ($1.26$).}}
  \caption{Overall, both designs were perceived as valuable, enjoyable, suitable and persuasive, with the RID slightly scoring better. The results are as follows [Mean (SD)]. \textit{RID}: Value 5.42 (0.95), Interest 6.18 (0.75), Suitability 6.2 (0.75), Persuasiveness 6.1 (1.04). \textit{SGED: Value 5.0 (0.45), Interest 5.95 (1.06), Suitability 5.3 (1.55), Persuasiveness 6.0 (1.26).} }~\label{fig:rid_sged_chart}
\end{figure}

The results of administering the IMI scales (\textit{Value / Usefulness} and \textit{Interest / Enjoyment}) and the statements regarding \textit{Suitability} and \textit{Persuasiveness} in 10 evaluation sessions on a 7-point likert scale (1 - 7) are shown in Figure \ref{fig:rid_sged_chart}. As depicted, the RID consistently scores better, although RID and SGED stimulate almost identical enjoyment and persuasion. In general, those two subscales score best, suggesting that our gamification designs are, overall, able to positively impact service contributions. The most pronounced difference between the two designs concerns the \textit{suitability} in research preservation. While the RID scores as well as in the previous subscales, the SGED is considered less suitable. Ordering effects with more than a one-point difference were observed only for \textit{SGED suitability} (SGED first: 6.4, SGED second: 4.2) and \textit{SGED persuasiveness} (SGED first: 6.8; SGED second: 5.2), suggesting that participants more critically reflected on controversial elements in the SGED, after having experienced the overall suitable RID. We focused on explaining this effect with the help of our extensive qualitative findings.

Overall, participants confirmed our underlying design assumptions, rating the following statements on a 5-point likert scale (Strongly disagree : -2\textbf{;} Strongly agree: 2):

\begin{itemize}
    \item I am willing to document and share my analyses as a contribution to the research quality of my collaboration: \textbf{1.8}
    \item I do \textbf{NOT} expect my career opportunities to profit from an increased visibility within my collaboration \textbf{( R )}: \textbf{-1.4}
    \item My analysis work would profit from access to the sum of well-documented <collaboration> analyses: \textbf{1.9}
    \item I would hope to engage in more collaboration, if I managed to increase my visibility within <collaboration>: \textbf{1.3}
    \item I think that I would want to frequently check the <collaboration> or group activities on the service dashboard: \textbf{0.9}
\end{itemize}

\section{Findings}

Four themes emerged from our qualitative data analysis. Here, we present each theme and our understanding of opportunities and constraints for designing gamification in science.

\subsection{Contribution}

Most participants (P1, P2, P5, P6, P8, P10) referred to improved career opportunities, resulting from game elements that reflect their contributions. To this end, a variety of mechanisms - from rankings to badges - seem valuable, as long as they can increase visibility within the huge collaborations:

\smallskip

\noindent \textit{"We are so many people in the collaborations, of course. Especially if we want to continue in our field, we have to get some visibility somehow."} (P6) \textit{"And if it's known that you were one of the first users of a particular technique, this can really help get your name out there."} (P2)

\smallskip

\noindent In this context, P1, P2, P6 and P10 explicitly mentioned their desire to refer to service achievements and scores in job applications. But the resulting competition also triggers concerns. In particular P2, P4, P6, P7 and P9 warn about unhealthy competition: \textit{"Imagine that my two PhD students had rank number 1 and rank number 2 and they compete with each other. I would find it a potential for a toxic situation."} (P7)

\subsubsection{Reflecting Contribution and Quality}

Given the potential impact of scores and achievements, all analysts discussed concerns related to accurately mapping research contributions in the gamification layer. P1 - P4 and P9 pointed to different roles within an analysis team. Concerning the preservation on the service, P3 notes that \textit{"it may be, for example, there is one person in a group who is taking care of this."} Thus, mechanisms are needed to reflect contributions:

\smallskip

\noindent \textit{"Maybe you can split the points for several people. Because if you give a big amount of points and only one person is allowed to push the button, this probably is a bit unfair. [...] You should find the means of splitting the awards or something."} (P1)

\smallskip

\noindent One physicist, P3, further worries that difficult tasks with low visibility might not be fully recognized, referring to the example of someone who struggles to solve an issue in programming code. P4 adds that metrics need to consider analysis complexity, because \textit{"if I preserve my shitty analysis 100 percent and someone else who actually was published in Nature preserves 60 percent, that does not really tell that my analysis is better than the other analysis."}

\subsubsection{Team Rather Than Individual Contributions}

Given the challenges that result from recognizing contributions, researchers (P1, P3, P7, P9, P10) strongly advocate promotion of team contributions, rather than personal ones. In fact, our analysis suggests that while competition on an individual level is controversial, comparison between teams and analyses is generally accepted.

\smallskip

\noindent \textit{"Any comparison between analyses, or everything you say about analyses I think it's very good. [...] I think people like to play that. But when you go inside one analysis things might get complicated."} (P9) \textit{"To boast that we do gracious things as a team. That would look less silly if it's at a team level. Rather than the individual that are gaining one more price."} (P7)

\subsection{Metrics}

A major theme that emerged from our data analysis relates to the selection of meaningful metrics in gamification design. Analysts described four dimensions: Frequency, accessibility, discouragement and social implications.

\subsubsection{Frequency}

A core dimension that has to be considered in the design of game elements is \textit{frequency} of contributions and activities. Most analysts referred to an expected unbalanced distribution of activities on the research preservation service. In particular P4 stressed that \textit{"it's just I feel there is this peak activity. People preserve in 3 days and then they stop."} Our data analysis revealed that the impact of frequency needs to be considered in various design elements. For example, both P2 and P4 commented on the SGED ribbon \textit{5th commit day}:

\smallskip

\noindent \textit{"So, I feel like there is a peak activity... that's why I feel that this 5th commit day is not so applicable." (P4)}

\smallskip

\noindent \textit{"Fifth means just like the fifth day in general, I think that's fine, but I would not want to encourage like continuous streaks of days, because I really don't like encouraging people to work outside their working hours. [...] And I also... At least when I work, I sort of dip in and out of projects quite frequently. So, I wouldn't want to have like any pressure to have to work on something for continuous block." (P2)}

\smallskip

P8 further depicts the effect of elements that are not frequently updated: \textit{"Then there are other things that stays, like yellow or red for you know a year. Then everyone just kind of stops paying attention. It turns to be more depressing than anything."}

\subsubsection{Accessibility}

The previous statement also highlights requirements connected to the accessibility of goals and achievements. Even though all participants acknowledged the analysis badges \textit{Executable} and \textit{Workflow Management Tool} to be important and valuable, analysts warned that the goals might be too high. P4 proposes to add more statuses to provide intermediate, accessible goals:

\smallskip

\noindent \textit{I think this maybe a too high goal to strive for. Because, I said that the biggest obstacle is probably that people know that it is going to take a lot of time. So, if you set them a very high standard... [...] it's like immediately judging them. It's not executable! So, I'm thinking maybe there should be more statuses. }(P4)

\smallskip

\noindent Both P7 and P8 referred to \textit{binary} mechanisms and highlighted that they are not likely to map reality. Concerning the analysis documentation progress bar (SGED) and pie chart (RID), P7 stated that \textit{"things are never binary. There is always partial completion. And one can think also about more than three categories."}

\subsubsection{Discouragement}
Participants highlighted adverse motivational effects resulting from discouraging statistics. Those are expected to be most pronounced in the early stages of the service's operation, where few activities and few preserved analyses are expected. Looking at the low overall documentation statistic of their collaboration on the RID version, P1 and P4 express their disappointment. P1, P4 and P7 propose to only show encouraging, positive information. For example:

\smallskip

\noindent \textit{"You want the star. [...] I guess it's an element that appears if you over-perform and does not appear otherwise."} (P7) \textit{"It's good for the preservation coordinator to show momentum. Of course, you would only show it if it's actually full."} (P1)

\subsubsection{Social implications}
Besides increasing visibility and improving career prospects, metrics also have social implications. In particular P2, P4, P5 and P9 comment on perceptions of the activity stream and collaboration documentation overview (RID). Looking at low numbers, P2 states: \textit{"If I saw that I'd be like: Maybe I can help get that number up."} The analysts describe their close identification with their collaboration. P4 even introduces the term \textit{tribalism}, to better illustrate the strong group feeling. Shown metrics can thus provoke social pressure:

\smallskip

\noindent \textit{"I think is cool, is to have the total goal for 2018, for instance. Like you really feel that you are contributing to the whole project, right."} (P5)

\smallskip

\noindent \textit{"... there are 20 people in this group. And then there is like higher probability that someone is going to make some activity. And then you are again going to feel like: oh my god, my peers are preserving. I should probably do the same thing."} (P4)

\subsection{Applications}

Our data analysis revealed that the gamification layer not only provides incentives and benefits to individual researchers; but instead can play an important role in two application areas: \textsc{Education / Training} and \textsc{Administration}.

\subsubsection{Education / Training}

Most analysts (P2, P3, P4, P7, P8, P9, P10) indicate that game elements can educate researchers about best practices. For example, P2 highlights \textit{"that an analysis that's like well documented, that's very reproducible, and does all the best practices, does probably end up with more views."} Thus, the researcher would like to sort by analysis views, to take inspiration from reproducible analyses. P2, P3, P4 and P8 highlight that those mechanisms can be most beneficial at the start of a new project and as learning material for new students. P8 even sees opportunities to impact the transition of practice:

\smallskip

\noindent \textit{"There is people who are doing things in an old way and then there is a new way of doing it where things are more reproducible etc. And what I think what we largely want to do is get signals to people of which ones are like doing it best practicy way and which ones aren't."}

\smallskip

\noindent In this context, P4 and P7 also caution about potential issues. P4 worries that the rank of analyses might not necessarily reflect suitability for teaching. Less complex analyses could score high, while more sophisticated ones might not. Yet, innovative, more complex analyses might set a better example.

Concerning the connection between point-based awards and elements that simulate best practices (SGED), P7 cautions about patronizing researchers. The convener also highlights that generally suitable practices might not apply in certain conditions in novelty-based research. Seeing the RID analysis page with the same workflow elements later, the convener judges the mechanisms to be suitable, because analysts are not forced to comply with a certain practice to get points.

\subsubsection{Administration}

Senior researchers (P1, P3, P7, P10) described how the transparency that is created by the gamification layer can be used in administrative tasks. The analysts indicate that the status transparency allows to more easily detect barriers. They describe detecting issues based on percentage-based documentation overview on the analysis level. In addition, P7 sees it as an opportunity to assess performance on a higher level:

\smallskip

\noindent \textit{"And maybe I can navigate in the front part. [...] To check who is over-performing or under-performing. To see what are the weak links and where to act. So, that's definitely the manager view and this sounds like the right thing to do in fact."}

\smallskip

\noindent In addition, P2 refers to the role of transparency and achievements in formal reviews. The analyst indicates that particularly the \textit{workflow management tool} and \textit{executable} badges would influence his perceived trust in the analysis.

\subsection{Scientific practice}

As our data analysis shows, the impact and requirements for game elements and mechanisms are manifold. Yet, a common denominator is the use of well-known scientific mechanisms.

\subsubsection{Speaking scientists' language}

Most participants (P3, P4, P6, P7, P8, P9) explicitly referred to the impact of design language on perceptions in a scientific environment. They highlight that design needs to adapt to serious, scientific language in order to be overall well-perceived: 

\noindent \textit{"It's probably for me - as a scientist... I'm disturbed, because it's sort of... I may be happy with gamification, but I don't want it to look like it."} (P3)

\smallskip

\noindent \textit{"The central part} (RID activity stream) \textit{"is professional. While the previous} (SGED leaderboards) \textit{"looks like something to engage a certain kind of people. [...] This is really professional and it's... Maybe it's less fun, but looks more useful."} (P7)

\smallskip

\noindent There is little controversy about game elements that use scientific language. While P3 considers community goals in collaboration statistics (SGED) to represent a \textit{"certain balance between the pure game type gamification elements and something which is sort of easily acceptable in a scientific domain"}, P4 argues that \textit{percentages} are already a strong and familiar metric for analysis completion, making points obsolete. P3 further highlights the strength of familiar language:

\smallskip

\noindent \textit{"Well, this gives sort of a scientific view of... Probably is more attractive to scientists because it gives you graphs. It's the language we speak, rather than points and awards and that kind of things. Which is something which is not our language in that sense. But it still gives you a scale."}

\subsubsection{Known mechanisms}

Besides familiar language, almost all analysts pointed to the suitability of known mechanisms. Concerning the analysis star in the RID, P3 and P4 see parallels to GitHub and GitLab mechanisms, commonly used code repositories. P1 instead compares achievement overviews and personal goals to mechanisms on Stackoverflow, a popular online developer community, and indicates that he would appreciate similar mechanisms in this context. P5 illustrates how points on the preservation service could potentially map to needed service points in real life. The researcher describes that analysts need to fulfill certain tasks as part of their obligations to the collaboration. Yet, \textit{"there are not so many opportunities to get this service points. And they are taken. So, somehow if you are able to arrive to some kind of agreement with the collaboration, for example CMS, and you can say like: I am going to change this many points in the analysis preservation page. I am going to exchange them by 1 service point."} Finally, P8 highlights the value of design elements that are more than just status elements, but rather provide a meaningful entry point:

\smallskip

\noindent \textit{"And also when you do that it gives you a little badge. Which says launch binder. Which in some sense is more like a button that looks like it does something. It's not just like collecting stars. It's an actionable something, you know. It also looks similar in terms of being you know a badge."}

\section{Discussion}
In the following, we discuss how our findings can be used to design engaging interactions through gamification in science. As our results suggest, a variety of game elements and mechanisms can provide value and enjoyment, while still being persuasive and suitable to the professional context. The overall low difference between RID and SGED in the quantitative assessment is not surprising, considering the work of Tondello, Mora and Nacke \cite{Tondello2017}. In their paper, they mapped 49 of the most frequently used gameful design elements, assigned to 8 groups, to gamification user types and personality traits and found that the overall difference \textit{"is not extraordinary but still pronounced, with approximately 20\% difference between the lowest and the highest scoring groups."} In addition, we see the overall low differences in our evaluation as evidence of the success of our extensive researcher-centered design process. Even though our qualitative findings highlight constraints and requirements of individual game mechanisms, researchers appreciated the underlying target behaviors and best practices that we aimed to stimulate, in turn creating overall acceptance.

In face of less pronounced quantitative differences, we consider the qualitative focus of our study to be a key strength. It allowed us to better understand the impact, opportunities and requirements of individual game mechanisms. In the following discussion, we see how prevalent challenges in scientific work need to be reflected in design requirements for gamification in science. We postulate that design has to consider \textit{controversial elements} very careful in this competitive environment. We conclude with \textit{design recommendations} and a note on how this relates to the success of open science badges \cite{kidwell2016badges}.

\subsection{Reflect the Scientific Environment and Contribution}

Scientists were particularly concerned about the reflection of research quality and personal contribution on the gamification layer. This means that designers \textbf{need to provide mechanisms that allow distributing awards and visibility justly mapping individuals' contributions.} This is a core requirement that applies to all game elements identifying efforts, but in particular to point-based rewards and rankings. In addition, it is important to \textbf{enable promotion of work based on quality, impact and purpose instead of relying solely on general, static service mechanisms.} This is particularly important as ranking and promotion of work has significant implications on education and training. Promoting work that does not fit these purposes risks providing misguided references for researchers that aim to learn about techniques or best practices. Thus, administrative or community mechanisms need to be created that allow to adapt ranking and visibility of work depending on the desired application.

Given multiple applications and uses of the gamification layer, systems should \textbf{allow adapting presentation to desired use cases}. In the preservation context, the system could provide filter mechanisms in analysis rankings tailored to training efforts, identifying work that implements best practices. As imagined also by one of the participants, presentation could be adapted based on the role of the user. Logging in to a system, senior researchers could profit from more visible performance overviews, while early-career researchers would most likely profit from awareness of relevant research activities.

Given that studies and analyses in science are often conducted over a long period of time, \textbf{it is crucial to provide accessible goals}. This applies particularly to research-related achievements. Awards that promote best practices should not only target the ultimate goal requiring months and years of effort, but intermediate steps. Whenever possible, binary reward mechanisms should be replaced by more multifaceted structures. Doing so is likely to prevent discouragement through facing a goal that is very hard to reach, but might instead provide a \textit{sense of progress}, one of the design pattern for gamification of work by Swacha and Muszy{\'{n}}ska \cite{Swacha2016}, making an \textit{"employee aware that every action he/she performs is a step in progress."} Yet, doing so might become more challenging in a scientific context, characterized by novelty and creativity. Our findings regarding the accessibility of goals also relates to research conducted on fitness trackers \cite{niess2018supporting}.

\subsection{Find Potential Breaking Points}

Our results suggest that both service designs are likely to be well-received in our setting, even though they made use of strongly contrasting gamification strategies. And even though most researchers generally approved of the designs, our qualitative analysis pointed to a fine line between particularly valuable and suitable design elements and those with a potential for controversy. Concerns were particularly pronounced for explicit personal rankings and point-based incentives, that by some participants were feared to patronize researchers and to limit them in their choice. Yet, others pointed to those mechanisms as their favorite design elements, allowing them to compete and aggressively promote necessary best practices. Given our findings, we consider those mechanisms to be highly \textit{controversial} and \textbf{system designers should weight potential costs and benefits employing controversial mechanisms}. 

Our findings suggest that independent of individual design elements, \textbf{mechanisms that promote team or analysis achievements are overall accepted, while personal promotion is controversial.} This can be seen particularly in statements referring to leaderboards and activity streams. While some researchers see personal metrics as a particularly strong opportunity to compete and to gain visibility, others worry about creating an unhealthy environment and a potentially \textit{toxic situation.} Yet promotion of collaborative achievements is overall accepted and desired, even if they employ the same design elements, namely leaderboards and activity streams.

There is little controversy regarding mechanisms and language known from established scientific practice. Our results suggest that \textbf{studying and integrating community-specific language profits perceived value and suitability}. Similarly, \textbf{the use of well-known mechanisms that are employed in common research tools} seems to be overall accepted.

\subsection{Role of Open Science Badges}

A recent systematic literature review concluded that open science badges are the only evidence-based incentive \cite{Rowhani-Farid2017} that promotes data sharing of research in the health and medical domain. In fact, in their quantitative study, Kidwell et al. \cite{kidwell2016badges} found a significant increase in data sharing of submissions to the \textit{Psychological Science} journal that adapted to those badges. Based on our findings and design implications, we discuss five aspects explaining why those mechanisms had a positive impact. First, the badges allow promoting best practices that are considered highly important in the community. We employed similar mechanisms in our study that were very well received by participants. Second, while badges are visibly placed on the paper and in the digital library of participating journals, \textit{no} adverse indication is given, highlighting that a paper has not yet received those awards. Third, promotion of rewarded papers increases their visibility, as well as the visibility of authors. This is especially true if search engines of digital libraries highlight corresponding search results. Through increased visibility, researchers can expect increasing citations and improved career prospects. Fourth, also the fact that badges are assigned to papers instead of individual researchers certainly fosters acceptance, as we have previously discussed. Finally, the badges provide accessible goals, a first step towards reproducibility. ACM takes this notion even further, introducing fine-grained badges that focus on very accessible goals \cite{boisvert2016incentivizing, acmbadgesweb}.

\subsection{Design Recommendations}

Our findings indicate that several dimensions need to be considered designing game elements in a research setting. Reflecting on previously discussed long periods required to fulfill research goals, \textbf{the design of game mechanisms should consider expected \textit{frequency} of status changes and activities}. Introducing intermediate, accessible goals - as we previously discussed - allows also to create activity for elements that are otherwise expected to stay in a certain condition for a long time, possibly frustrating researchers that do not see an opportunity to communicate progress. At the same time, designing elements with expected impact on frequency of use also allows preventing unwanted mechanisms, in particular forcing scientists to continuously perform actions, even though such streak mechanisms might not be suitable to the context. 

Also related to \textit{frequency}, our findings suggest that \textbf{design needs to deal with potentially \textit{discouraging} statistics and messages.} This concerns both collaboration-wide statistics as well as elements that depict the status of analyses. Particularly in the early stages of a service or analysis. In response, messages can elaborate why statistics are less promising than most would expect, pointing for example to the fact that a service became operational only a short while ago. Related to the presentation of work, some elements - for example the popularity star or a badge - might only appear in case it is earned, instead of illustrating that the reward can still be earned.

Finally, we encourage to systematically \textbf{consider social factors resulting from design}. Not only have our findings shown that researchers agree on mechanisms that foster cooperation and that they find individual promotion controversial; we also perceived indications of positive social pressure. Statistics and elements that depict activity are likely going to create peer pressure, in particular if researchers have a strong identification with their research collaboration. 

\section{Limitations and Future Work}

The findings and design implications of our study are solely based on evaluations with HEP researchers. We discussed how they relate to the underlying mechanisms of Open Science Badges. Those represent a strong example of successful game elements in (reproducible) science, used in a variety of scientific fields. Thus, we anticipate that the design implications presented may be applicable in other fields of research. We envision potential for future research, mapping requirements of gamification in diverse scientific domains to gather additional requirements resulting from differing practices and needs. To foster future work, we release relevant resources, in particular the interactive Balsamiq archives, the questionnaires, questionnaire responses and the semi-structure interview guide. Based on our findings, we also envision further research studying the effects of controversial design mechanisms, in particular related to personality types of scientists. This might allow providing individual, personality-based experiences that further mitigate use of alienating mechanisms for some researchers, while providing stimulating ones to others.

The rudimentary nature of the prototypes and the lack of deployment represent both a limitation to this study, as well as a necessary step in the systematic study of requirements for gamification in a highly skilled scientific environment. The prototypes allowed confronting researchers with a wide variety of very different game elements, without the risk of deploying a design that may have negative consequences. We also consider the number of participants suitable for the qualitative focus of this study. The relatively high ratings of the questionnaire concerning Value, Interest, Suitability and Persuasiveness represent a valuable indicator for the potential of gamification in science. Yet, we recognize that the information value of the questionnaire would profit from higher participant numbers. In this context, we further envision implementation and evaluation of our design implications in production research tools. Even though we have perceived the interactive prototypes to be suitable for evaluation with our participants, production systems would allow mapping researchers' behaviors and perceptions over a longer time, interacting with real colleagues. That way, data from more researchers could be collected without requiring a strenuous recruitment process for volunteer participation of highly skilled and busy individuals.

\section{Conclusion}

This paper presented a systematic study of perceptions and requirements of gamification in science. We conducted our study in the context of research reproducibility, one of the most prevalent challenges in science today that suffers from motivating researchers to document and preserve their work; repetitive and often unrewarding tasks. Through several research activities, we learned about opportunities in designing gamification for research preservation in data-intensive experimental physics. Based on our researcher-centered design, we created two interactive prototypes of a preservation service that make use of contrasting gamification strategies.

Our evaluation showed that both the rational-informative as well as the openly competitive designs were considered valuable, enjoyable, suitable and persuasive. Through thematic analysis of our interviews, we identified four themes to inform about design requirements for gamification in science: Contribution, metrics, applications and scientific practice. Our data analysis revealed that gamification needs to address well-known challenges of the research process, including the fair reflection of quality and individual contribution. Our findings point to a high level of controversy related to the promotion of individual achievements and suggest that team and analysis-related rewards are generally accepted and desired.

Finally, we discussed implications designing for gamification in science that we expect to impact prevalent scientific challenges. We further discussed how already implemented open science badges relate to our design implications.

\balance{}

\balance{}

\begin{acks}
This work has been sponsored by the Wolfgang Gentner Programme of the German Federal Ministry of Education and Research (grant no. 05E15CHA).
\end{acks}

\bibliographystyle{ACM-Reference-Format}
\bibliography{sigchi}

\end{document}